\RequirePackage[2020-02-02]{latexrelease}
\documentclass[prd,nofootinbib,12pt]{revtex4}
\usepackage{amssymb,amsmath,latexsym,braket,fancyref,hyphenat}

\usepackage{graphicx}% Include figure files
\usepackage{dcolumn}% Align table columns on decimal point
\usepackage{bm}% bold math
\usepackage{placeins}

\newcommand{\hf}{\frac{1}{2}}

\newcommand{\ba}{\begin{align}}
\newcommand{\ea}{\end{align}}

\numberwithin{equation}{section}
\usepackage{parskip}
\begin{document}

%\begin{flushright}
%MAN/HEP/2022-xxx\\
%June 2022
%\end{flushright}

\title{{\Large Numerical Estimations of the Distribution of the Lifetime of Bubbles Emerging from First Order Cosmological Phase Transitions}\\[3mm] }

\author{Mulham Hijazi$^{\,a}\,$\footnote{E-mail address: {\tt mulham.hijazi@postgrad.manchester.ac.uk}}}

\affiliation{\vspace{2mm}${}^a$Department of Physics and Astronomy,
  University of Manchester}

\begin{abstract}
\vspace{2mm}
\centerline{\small {\bf ABSTRACT} }
\vspace{2mm}

We present a mathematical framework to produce a numerical estimation to the distribution of the lifetime of bubbles emerging from first order cosmological phase transitions. In a precedent work, we have implemented the Sound Shell model to predict the power spectra of gravitational waves arising from the decay of scalar fields. The model depends on the lifetime distribution of bubbles before collision, which in turn depends on the transition rate $\beta$ and the speed of the bubble wall $v$. Empirical exponential laws were used to describe the lifetime distribution and the resultant power spectra. For detonations, the results show a good agreement with simulations where the bubbles have nucleated simultaneously with a mean separation distance. However, for deflagrations, the results show that the amplitude of gravitational waves is higher at longer wavelength than simultaneous nucleation, indicating the importance of having a more accurate description of the lifetime distribution of bubble lifetime.

\end{abstract}

\maketitle

\vspace{10mm}
\section{Introduction}

The standard model of particle physics predicts that massive particles gain their mass through the Higgs mechanism, as the value of their masses  is proportional to the vacuum expectation value of the scalar field \cite{Goldstone:1962es,Gleiser:1998kk,Bailin:2004zd,Linde:1978px}. Thermal field theory predicts that the potential of the scalar field is affected by the temperature of the Universe. The effective potential is of the form:

\begin{align}
V_\text{eff}(\phi,T)= \frac{D}{2}(T^2-T_0^2)|\phi|^2-\frac{E}{3}T|\phi|^3+\frac{\lambda}{4}|\phi|^4,
\end{align}
where $D$, $T_0$, $E$, and $\lambda$ are constants \cite{Laine:2016hma,Linde:1983px,Vainshtein:1981wh,Linde:1981zj}. We can see that above the critical temperature, $T_c=T_0/\sqrt{1-2E^2/9\lambda D}$, the potential becomes symmetric and the minimum of the potential occurs only at $\phi=0$. Thus, we expect that standard model particles were massless at early times of the history of the Universe and the fundamental forces of nature were unified. 

However, as the Universe cools down below the critical temperature, the potential shapes up to have a second minima, a true vacuum of the theory. Quantum mechanical laws allow for the Universe to tunnel through the potential barrier to reside in the true vacuum, as energetics favour the tunneling to the new lowest ground state of the vacuum.

\begin{figure}[h!]
\begin{center}
\includegraphics[width=.75\textwidth]{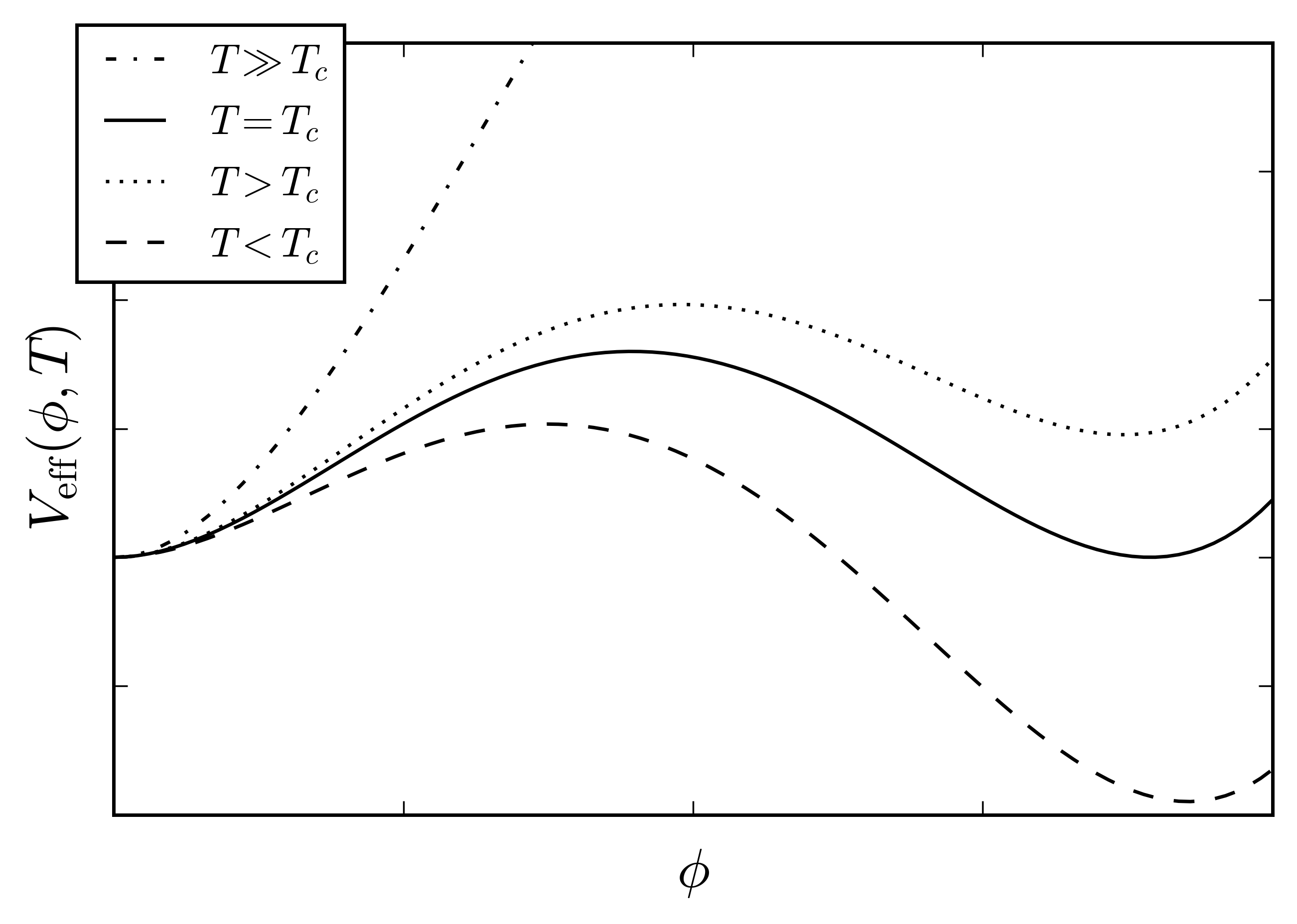}
\caption[Effective Higgs potential $V_\text{eff}(\phi,T)$ at finite temperatures]{Effective Higgs field potential $V_\text{eff}(\phi,T)$ at finite temperatures}
%\label{fig: myfig}
\end{center}
\end{figure}

The mathematical foundations describing the process of vacuum decay were first laid out by Coleman, where he introduced the so-called the thin wall solution to the equation of motion corresponding to a potential where the difference between the false and true vacuum is small compared to the height of the barrier between them. The tunneling rate per unit volume is given by \cite{Coleman:1977py,Coleman:1978ae,Rajaraman:1982is,Shifman:1994ee,Rubakov:1984bh,Callan:1977pt,Coleman:1980aw}
\begin{align}
\Gamma/V=A e^{-B},
\end{align}
where $A$ is a prefactor, and the exponent $B$ is the difference between the Euclidean action $S_E$ of the bounce and false vacuum solutions. The mechanism of which vacuum decay occurs is through the nucleation of bubbles which grow and fill up the entire spacetime continuum, with the interior of these bubbles residing in the true vacuum. The decay of the vacuum leads to cosmological phase transitions as standard model particles gain mass, breaking gauge invariances in the process. Moreover, studying the stability of the vacuum helps us set constraints on physical constants in particle physics models \cite{Branchina:2018xdh,  Linde:2007fr,Frampton:1976pb,Weinberg:2012pjx,Hijazi:2019fme}. Signatures of such decays may manifest themselves in the form of resultant gravitational waves which we aim to detect in the future. Thus, several papers have aimed to predict the shape of the power spectra of these gravitational waves \cite{Hindmarsh:2015qjv,Kamionkowski:1993fg,Jinno:2016vai,Weir:2016tov,Hindmarsh:2017gnf}.  

In a precedent work, we have implemented the Sound shell model to predict the shape of the power spectra of the gravitational waves resulting from these cosmological phase transitions \cite{Hindmarsh:2016lnk,Hindmarsh:2019phv}. These gravitational waves are sourced by the explosive growth of bubbles of the true vacuum, governed by the hydrodynamics occurring at the bubble walls. The velocity profiles of the cosmological fluids surrounding the bubbles are classified into detonantions, deflagrations, and hybrids \cite{Espinosa:2010hh,Sopena:2010zz,Ignatius:1993qn}. 

The shape of the power spectra is affected by the lifetime distribution of these bubbles before they collide. An empirical exponential law was used to describe this distribution and the results were compared to the power spectra predicted from simulations where simultaneous nucleation of bubbles with a fixed separation distance was assumed. The results were in good agreement for detonations, however for deflagrations the results differed, showing a higher amplitude at longer wavelengths \cite{Hindmarsh:2019phv}, indicating that a more accurate description of the lifetime of the vacuum is needed. Although it was argued in another paper that the source of the discrepency in results for deflagrations was due to the reduction of kenitic energy due to the integration of the sound shells \cite{Cutting:2019zws}.

In this paper, we present a mathematical framework in order to numerically estimate the distribution of the lifetime of bubbles before their collision. In Section II, we will lay out the theoretical background needed to understand the nature of the problem, and then proceed to derive mathematical expressions to estimate the lifetime of nucleating bubbles. In Section III, we will present our results for a range of values for the transition rate $\beta$ and bubble wall speeds $v$, and compare our results to bubble lifetime distributions resulting from simulations of randomly generated periodic Universes described by a unit cell with a fixed number of bubbles. At last, Section IV will discuss the results and summarise our conclusions.

\section{Theoretical Background}\label{sec:Theory}

The difficulty in calculating an exact distribution of the lifetime of bubbles lies in the fact that vacuum decay is a probabilistic process, which entails that these bubbles nucleate at random locations and times. Moreover, bubbles only nucleate in the metastable phase which means that the space available for these bubbles to nucleate shrinks over time as more bubbles nucleate and grow and eventually fill up the entire space. 

However, an estimate to the number of bubbles nucleated $N(t)$, and the fraction of space which resides in the false vacuum $h(t)$, as a function of time was worked out by Enqvist \textit{et al}\cite{Enqvist:1991xw}, as we quote their results:
\begin{align}
h(t)=\exp(-e^{\beta(t-t_f)}),\\
N(t)=V\frac{\beta^3}{8\pi v^3} (1-h(t)),
\end{align}
where $\beta$ is the transition rate, $v$ is the bubble wall velocity, and $t_f$ is some arbitrary time chosen such that $h(t_f)=1/e$. The value of $t_f$ is irrelevant to our discussion since we are only interested in time differences between the time of nucleation and the time of collision.

The function $h(t)$ is an exponentially decaying function with the property that $h(\infty)=0$ as bubbles grow to fill the entire spacetime continuum. Hence, the total number of bubbles nucleated at the end of the phase transition is given by
\begin{align}
N_{\text{tot}}\equiv N(\infty)=V\frac{\beta^3}{8\pi v^3}.	
\end{align} 
In our analysis we will discretise the time of nucleation of each bubble as 
\begin{align}
\label{eq:tn}
 t_n\equiv\frac{1}{\beta}\ln\bigg(-\ln\bigg(1-\frac{n}{N_{\text{tot}}}\bigg)\bigg) +t_f,
\end{align}
where $t_n$ is the time of nucleation of the $n$th bubble. We then define the function $R_n(t)$ describing  the radius of the $n$th bubble as a function of time as
\begin{align}
R_n(t)=v(t-t_n),
\end{align}
where $t>t_n$. We denote the time of which the bubble, say $b_i$, has its first collision with another bubble, say $b_j$, as $t_*$. For this to occur, two conditions should be met:

1- Any bubble $b_k$ ($k\neq j$)  nucleated at time $t_k <t_*$ should nucleate at a distance from $b_i$ greater than the sum of their radii at $t_*$, otherwise $b_k$ would collide with $b_i$ before $b_j$. This implies that for each bubble $b_k$ there is a corresponding ``Forbidden volume'' of the metastable phase for which the bubble could not have nucleated within. For bubbles which have nucleated before $b_i$, $t_k<t_i$, the forbidden volume is described by a sphere with a radius defined by the minimum allowed distance between them $R_i(t_*)+R_k(t_*)$. For bubbles which have nucleated after $b_i$, $t_k>t_i$, the forbidden volume is given by the volume of the sphere enclosed by the minimum allowed distance subtracted by the volume of the bubble $b_i$ at $t_k$, since the volume of $b_i$ at that time resides in the true vacuum.

2- The bubble $b_j$ should nucleate on the surface of the sphere parameterised by $R(t_*)\equiv R_i(t_*)+R_j(t_*)$, with $dR=2vdt_*$.

\begin{figure}[ht!]
\begin{center}
\includegraphics[width=0.75\textwidth]{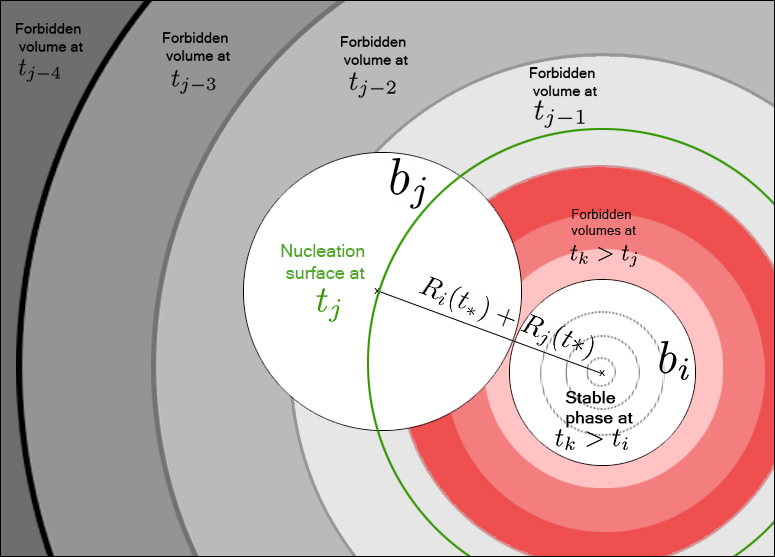}
\caption{Schematic diagram of the cross section of bubble $b_i$ as it collides for the first time with bubble $b_j$ at time $t_*$. The grey and red areas represent the forbidden volume where the corresponding bubbles could not have nucleated within, with the darker shades referring to earlier times. The green perimeter represents the surface of which bubble $b_j$ must nucleate at in order to collide with $b_i$ at time $t_*$ exactly. The dashed circles represent the size of the bubble $b_i$ at different nucleation times $t_k>t_i$ which we need to subtract from the forbidden volumes since the interior of the bubble resides in the stable phase.}
\label{fig:diag}
\end{center}
\end{figure}

From that we infer that the probability that $b_j$ is the first bubble to collide with $b_i$, and that the collision occurs at $t_*$  is given by
\begin{align}
\label{eq:prob}
P_{ij}(t_*)dt_*=N_i& \prod_{k=1,k\neq j}^{k=i-1} \bigg(1-\frac{\frac{4\pi}{3} (R_i(t_*)+R_k(t_*))^3}{h(t_k)V}\bigg)\nonumber\\
& \prod_{k=i+1,k\neq j}^{k=m} \bigg(1-\frac{\frac{4\pi}{3} [(R_i(t_*)+R_k(t_*))^3-R_i(t_k)^3]}{h(t_k)V}\bigg)\nonumber\\
&\times \frac{8\pi R^2(t_*)vdt_*}{h(t_j)V}    \times \theta[h(t_j)V-\frac{4\pi}{3} R^3(t_*)],
\end{align}
where $m$ is defined such that $t_m = \text{max}\{t_i; t_i<t_*\}$.  The first two lines of \eqref{eq:prob} correspond to the product of the probability that all bubbles $b_k$ ($k\neq j, t_k<t_*$) have nucleated in their corresponding allowed regions of space. The last line gives the probability that the bubble $b_j$ have nucleated on the surface defined by a sphere enclosed by $R$. The $\theta$ function ensures that the probability of a (first) collision happening vanishes if the sphere enclosed by the distance between the two bubbles is larger than the volume of the false vacuum at $t_j$. The normalisation constant $N_i$ is fixed by the condition 
\begin{align}
\sum_j \int^\infty_{\text{max}\{t_i,t_j\}} P_{ij}(t_*)dt_*=1.
\end{align}
\begin{figure}[ht!]
\begin{center}
\includegraphics[width=0.75\textwidth]{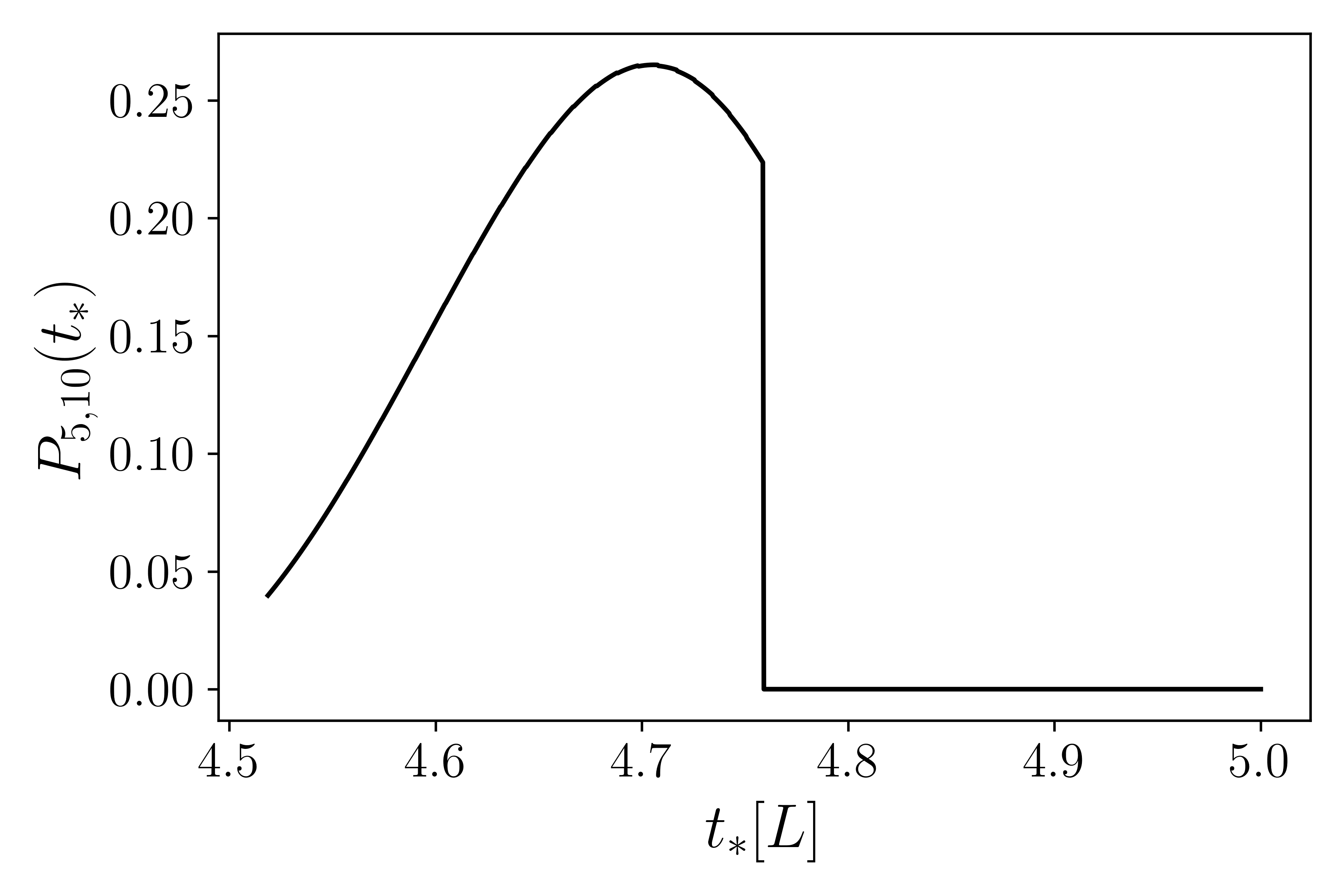}
\caption{The weighted probability density that bubble $b_5$ had its first collision with bubble $b_{10}$ as a function of collision time $t_*$, for values $\beta=5 \ L^{-1}, v=0.35$, $V=1 \ {L}^3$  and $t_f=5 \ L$. The probability plummets at the time $t_*$ when the sphere enclosed by the radius $R_5(t_*)+R_{10}(t_*)$ is larger than the volume of the metastable phase $h(t_*)$.}
\label{fig:diag}
\end{center}
\end{figure}
Now we can write down an expression for the average lifetime $\bar{T}_i$ of the bubble $b_i$ as
\begin{align}
\bar{T}_i= \bigg[ \sum_j \int^\infty_{\text{max}\{t_i,t_j\}} t_* P_{ij}(t_*)dt_* \bigg]  - t_i.
\end{align}
Finally, after computing the lifetime of each bubble, we can fit a distribution using the histogram of lifetimes of all bubbles. In the next section we will numerically find the lifetime distribution of bubbles corresponding to a range of values for the transition rate $\beta$ and various bubble wall speeds $v$.

\section{Results}

In our analysis, we consider a unit volume $V= 1 \ L^3$, and fix the value of $t_f= 5 \ L$ arbitrarily. We run our computations for transition rate ranging between $\beta = 5-10 \ L^{-1}$, bubble wall speeds ranging between $v=0.25-0.5 $, and number of bubbles in a unit volume within the range $N_{\text{tot}}=54-318$. 

We proceed to numerically fit the distribution of bubble lifetimes $\nu_{T}(\bar{T}_i)$ as a function of the average life time $\bar{T}_i$ for each bubble. We compute the average lifetime for all bubbles $\bar{T}$ and the standard deviation $\sigma_{\bar{T}}$.  We then define
\begin{align}
\bar{R}_i= v\bar{T}_i,
\end{align}
as the average radius of bubble $b_i$ when it collides for the first time. Consequently we find the average radius at the time of collision for all bubbles $\bar{R}=v\bar{T}$, and the standard deviation $\sigma_{\bar{R}}= v\sigma_{\bar{T}}$.
Furthermore, we define 
\begin{align}
R_{\text{uni}} =\hf n^{-1/3}_{\text{tot}} \equiv  \hf \bigg(\frac{N_{\text{tot}}}{V}\bigg)^{-1/3}
\end{align}
as the radius of bubbles which have nucleated simultaneously and uniformly at the time of their collision. This expression was used in simulations where gravitational waves were generated from simultaneous nucleation of bubbles \cite{Hindmarsh:2015qjv}.

\begin{table}[!ht]
\setlength{\tabcolsep}{10pt} % Default value: 6pt
\renewcommand{\arraystretch}{1.5} % Default value: 1
\centering
\begin{tabular}{cccccccc}
\hline
$\beta[\text{L}^{-1}]$&$v$&$\lfloor N_{\text{tot}}\rfloor$ &$ \bar{T}[\text{L}]$&$\sigma_{\bar{T}}[\text{L}]$&$\bar{R}[\text{L}]$&$\sigma_{\bar{R}}[\text{L}]$& $R_{\text{uni}}[\text{L}]$\\
\hline
5&0.25&318&0.0994&0.0832&0.0249&0.0208&0.0732\\
5&0.35&116&0.0878&0.0758&0.0307&0.0265&0.1025\\
5&0.45&54&0.0724&0.0680&0.0326&0.0306&0.1318\\
6&0.35&200&0.0799&0.0669&0.0280&0.0234&0.0854\\
7&0.35&318&0.0708&0.0595&0.0248&0.0208&0.0732\\
10&0.5&318&0.0494&0.0417&0.0247&0.0209&0.0732\\
\hline
\end{tabular}
\caption{Numerical estimates of the average lifetime $\bar{T}$, the standard deviation $\sigma_{\bar{T}}$, the corresponding average radius of the bubble at the time of its first collision $\bar{R}$, the standard deviation $\sigma_{\bar{R}}$, and the average radius of bubbles at the time of their collision when they are nucleated simultaneously and uniformly $R_\text{uni}$ for different input values of the the transition rate $\beta$ and bubble wall speeds $v$.}
\label{tab:Results}
\end{table}

\begin{figure}[ht!]
\begin{center}
\includegraphics[width=0.75\textwidth]{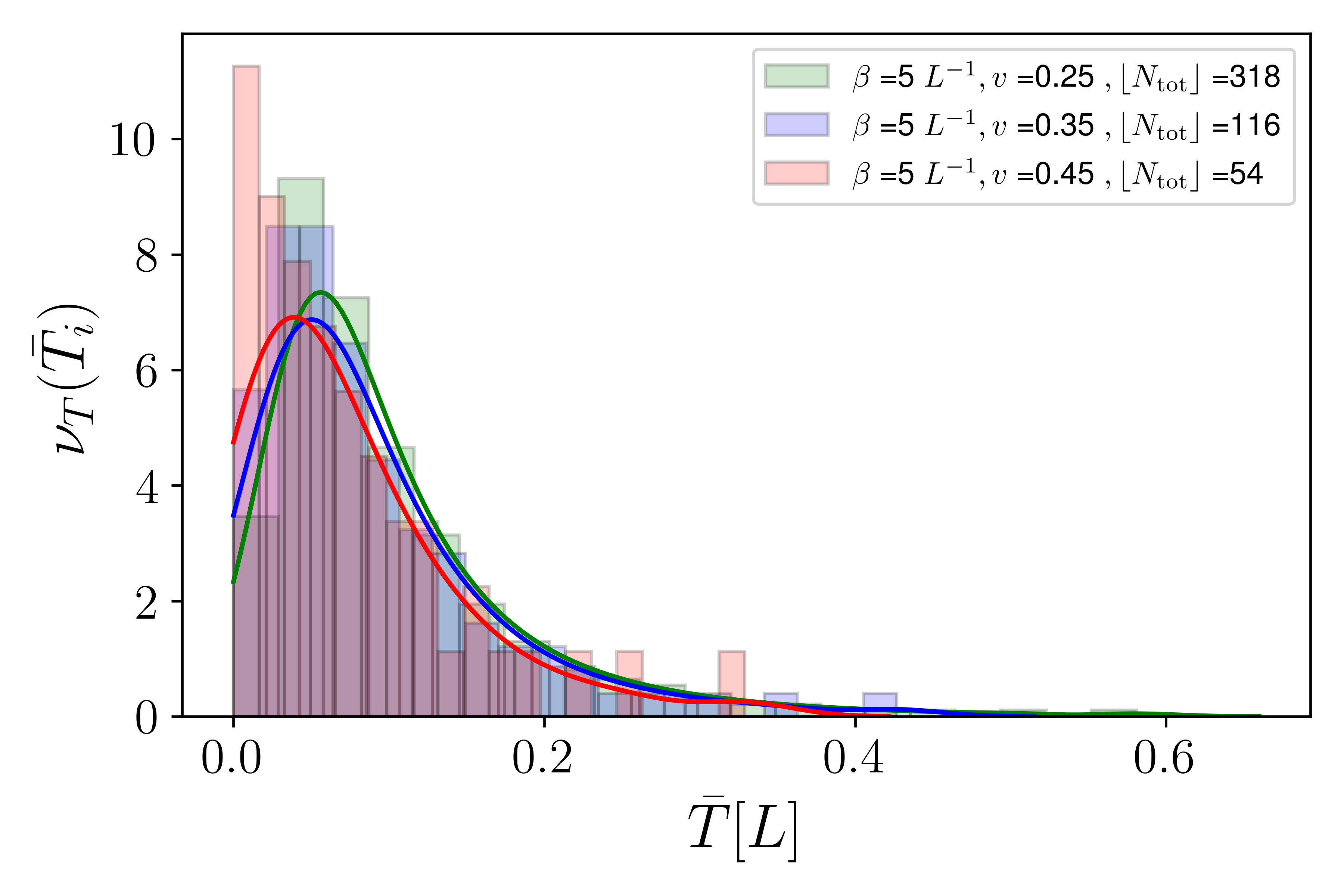}
\caption{The lifetime distribution $\nu_{T}(\bar{T}_i)$ for a fixed value for the transition rate $\beta=5 \ L^{-1}$, and varying values for the number of bubbles in a unit volume $\lfloor N_{\text{tot}}\rfloor$, and bubble wall speeds $v$.}
%\caption{Schematic diagram of the cross section of bubble $b_i$ as it collides for the first time with non other than bubble $b_j$ at time $t_*$. The grey and red areas represent the forbidden volume where the corresponding bubbles could not have nucleated within, with the darker shades referring to earlier times. The green perimeter represents the surface of which bubble $b_j$ must nucleate at in order to collide with $b_i$ at time $t_*$ exactly. The dashed circles represent the size of the bubble $b_i$ at different nucleation times $t_k>t_i$ which we need to subtract from the forbidden volume since the interior of the bubble resides in the stable phase.}
\label{fig:b}
\end{center}
\end{figure}

\begin{figure}[ht!]
\begin{center}
\includegraphics[width=0.75\textwidth]{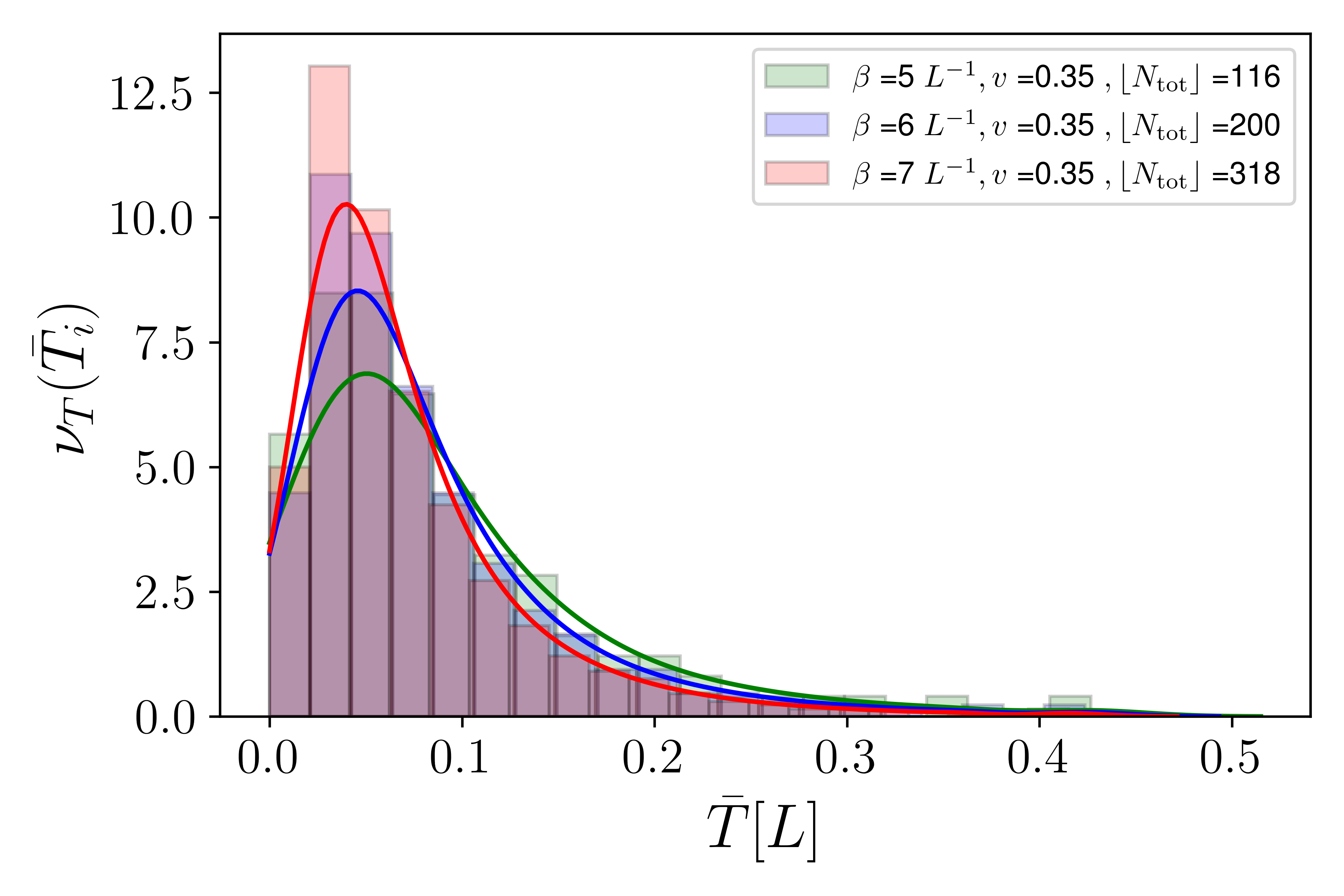}
\caption{The lifetime distribution $\nu_{T}(\bar{T}_i)$ for a fixed value for the bubble wall speed $v=0.35 $, and varying values for the number of bubbles in a unit volume $\lfloor N_{\text{tot}}\rfloor$, and transition rate $\beta$.}
%\caption{Schematic diagram of the cross section of bubble $b_i$ as it collides for the first time with non other than bubble $b_j$ at time $t_*$. The grey and red areas represent the forbidden volume where the corresponding bubbles could not have nucleated within, with the darker shades referring to earlier times. The green perimeter represents the surface of which bubble $b_j$ must nucleate at in order to collide with $b_i$ at time $t_*$ exactly. The dashed circles represent the size of the bubble $b_i$ at different nucleation times $t_k>t_i$ which we need to subtract from the forbidden volume since the interior of the bubble resides in the stable phase.}
\label{fig:v}
\end{center}
\end{figure}

\begin{figure}[ht!]
\begin{center}
\includegraphics[width=0.75\textwidth]{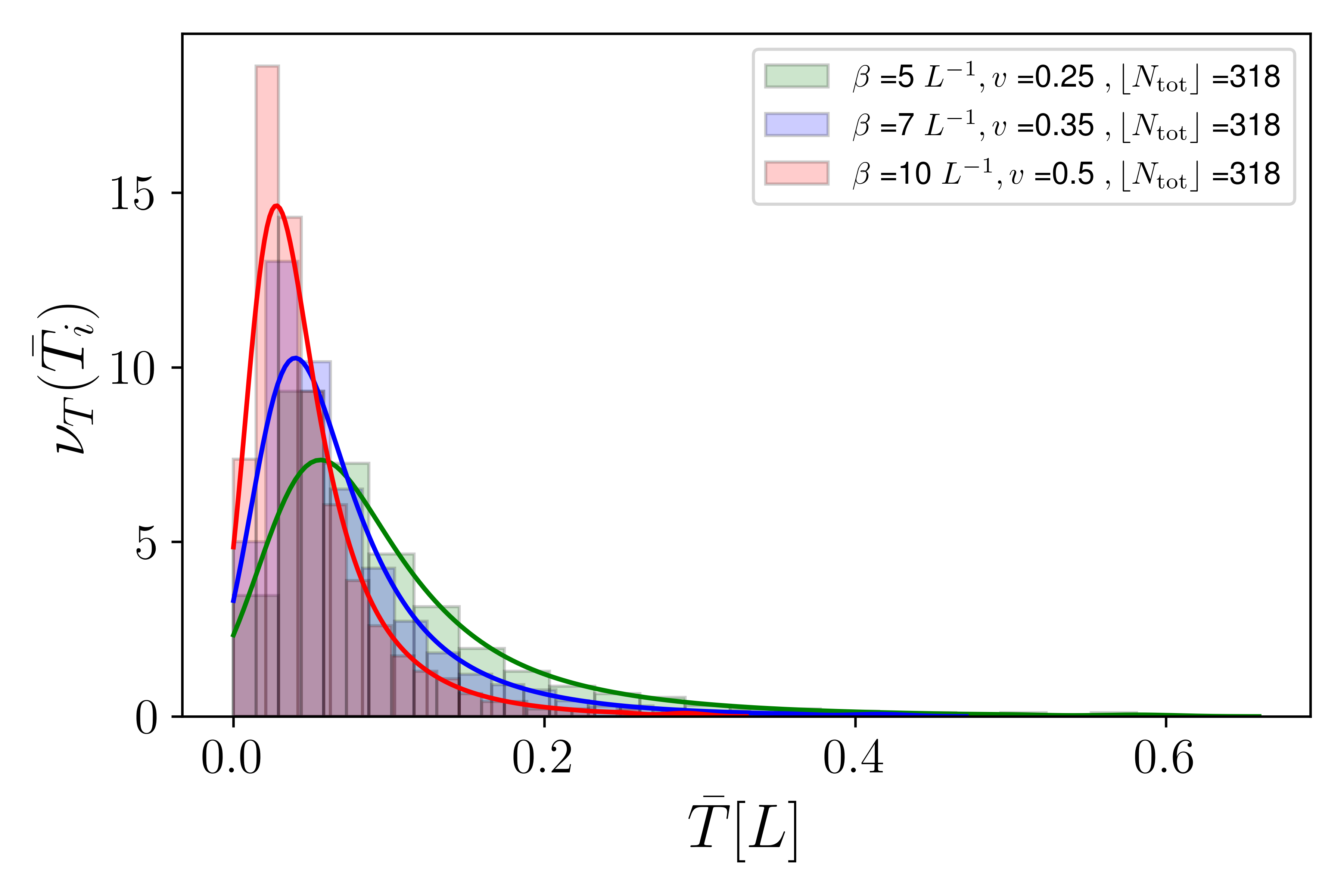}
\caption{The lifetime distribution $\nu_{T}(\bar{T}_i)$ for a fixed number of bubbles in a unit volume $\lfloor N_{\text{tot}}\rfloor =318$, and varying values for the transition rate $\beta$, and bubble wall speeds $v$.}
\label{fig:Nb}
\end{center}
\end{figure}

The numerics are laid out in Table ~\ref{tab:Results}, and plots of the lifetime distribution $\nu_{T}(\bar{T}_i)$ are displayed in Figures~\ref{fig:b},~\ref{fig:v},~\ref{fig:Nb}. We notice that the average lifetime $\bar{T}$ and the standard deviation $\sigma_{\bar{T}}$ decrease as the transition rate $\beta$ increases. They also decrease as bubble wall speeds $v$ increase.  
\\
\\

Moreover, The average radius of the bubble at the time of collision $\bar{R}$ is smaller than the radius of bubbles which have nucleated simultaneously and uniformly at the time of collision $R_{\text{uni}}$. This was an expected result since bubbles which have nucleated at later times can only nucleate within small pockets of space that still reside in the false vacuum.

\begin{figure}[ht!]
\begin{center}
\includegraphics[width=0.75\textwidth]{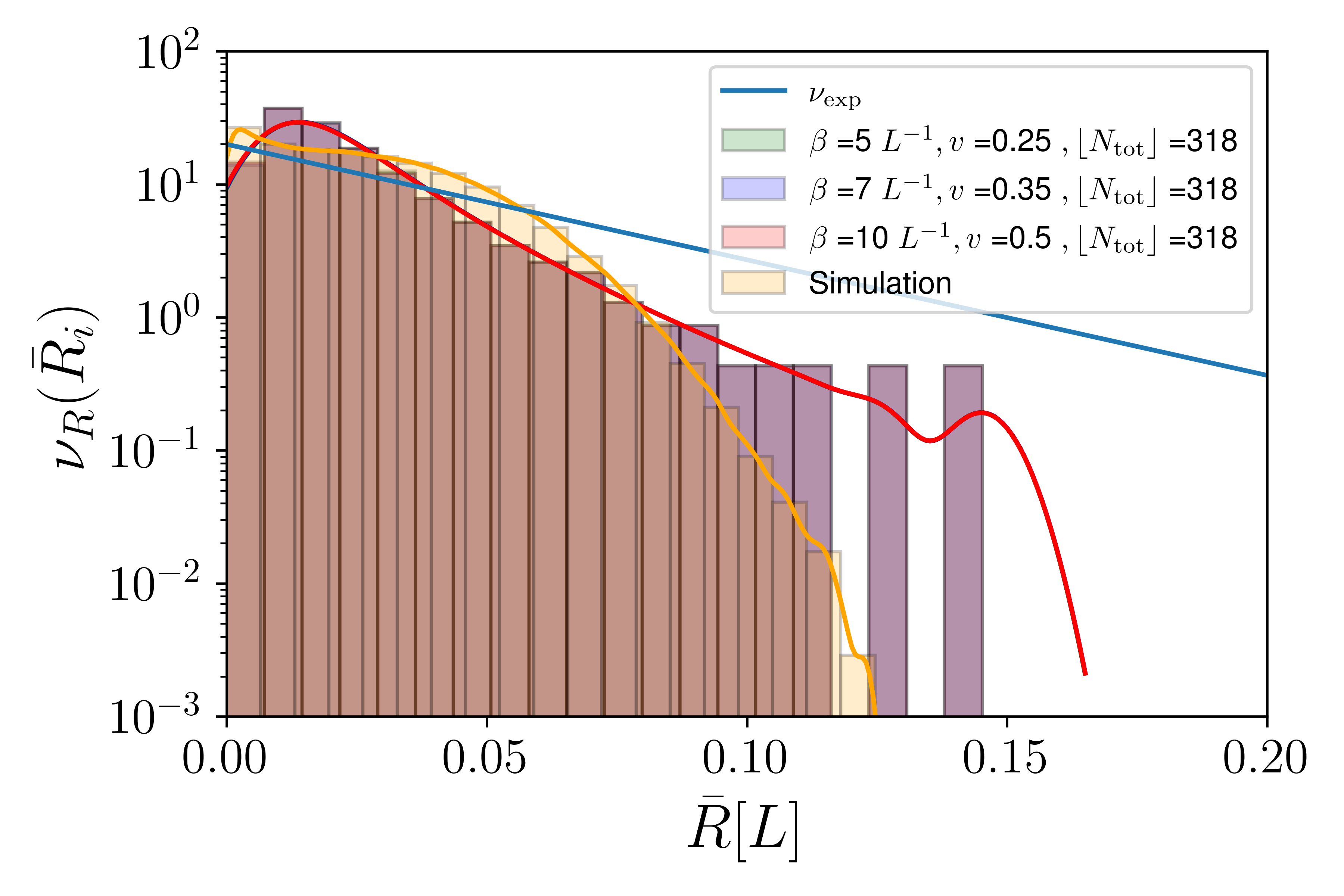}
\caption{The bubble size distribution $\nu_{R}(\bar{R},\sigma_{\bar{R}})$ for a fixed number of bubbles in a unit volume $\lfloor N_{\text{tot}}\rfloor =318$, and varying values for the transition rate $\beta$, and bubble wall speeds $v$. We notice that the size distributions fit identically for decays where the same number of bubbles are nucleated. The bubble size distribution shown in yellow results from stacking up 1000 simulations of randomly generated periodic Universes with the same number of bubbles per unit volume $n_{\text{tot}}=318$ where the times of nucleation are described by $t_n$, fixing the value of the transition rate at $\beta =5 \  L^{-1}$, and bubble wall speed at $v=0.25$. The distribution given in blue describes the size distribution resulting from exponential nucleation quoted in Hindmarsh \textit{et al} for the same values for the physical parameters given in the simulations.}
\label{fig:nur}
\end{center}
\end{figure}

Interestingly, we notice that the values computed for the average radius of the bubble at the time of collision $\bar{R}$, and the standard deviation $\sigma_{\bar{R}}$ corresponding to different input parameters for the transition rate $\beta$ and bubble wall speeds $v$ which yield the same total number of bubbles $N_{\text{tot}}$ are roughly the same. This might indicate that the distribution of bubble sizes at the time of collision $\nu_R(\bar{R}_i)$, at least for some range of parameters, depends only the number of bubbles $N_{\text{tot}}$ as shown in Figure~\ref{fig:nur}.  

In addition, we have stacked up 1000 simulations where we have generated periodic Universes described by a unit cube $L^3$ where we have placed $N_{\text{tot}}=318$ nucleation points randomly within the cube on the condition that all bubbles nucleate in the metastable phase. The time of nucleation of the $n$th bubble $t_n$ is given by~\eqref{eq:tn}. We fixed the values of the transition rate at $\beta=5 \  L^{-1}$, and the wall speed at $v=0.25$. Then we proceeded to compute the radius of each bubble at the time of its first collision.  The average radius of bubbles at the the time of their first collision is $\bar{R}=0.0269 \  L$, which is very close to the values shown in Table ~\ref{tab:Results} for transitions which yield $\lfloor N_{\text{tot}}\rfloor=318$. 

We created a histogram of these radii and plotted it against the histogram obtained from our mathematical framework in Figure~\ref{fig:nur}. The distributions are in good agreement as they have roughly the same average and roughly the same shape, although the peak of the distribution in the simulations seems to be at a slightly smaller size which is likely due to the periodicity of the Universe in the simulations. Furthermore, we plotted the bubble size distribution given in \cite{Hindmarsh:2019phv} for exponential nucleations, described by

\begin{align}
\nu_{\text{exp}}(R)= \frac{\beta}{v} e^{-\beta/v R},
\end{align} 

for the same values for the physical parameters given in the simulations. We can clearly see that it fails to replicate the distributions that resulted from the simulations at larger radii. The average radius of bubbles is also much larger as the distribution yields the average $\bar{R} = 0.05 L$.

\section{Conclusion}

We presented a mathematical framework to compute an estimate to the lifetime distribution of bubbles  $\nu_{T}(\bar{T}_i)$ as a function of the transition rate $\beta$, and bubble walls speed $v$. We started by discretising the time of nucleation of each bubble $t_n$ and then calculating the probability that bubble $b_i$ had its first collision with bubble $b_j$ at time $t_*$.  This was done by calculating the probability that every other bubble had nucleated outside the forbidden region of the metastable phase, and that bubble $b_j$ has nucleated on the surface parameterised by the radius $R_i(t_*)+R_j(t_*)$.

After computing the average lifetime of each bubble $\bar{T}_i$, we create a histogram and fit the distribution of bubble lifetimes   $\nu_{T}(\bar{T}_i)$. We ran our computations using different values for the transition rates ranging between $\beta = 5-10 \ L^{-1}$, bubble wall speeds ranging between $v=0.25-0.5 \ $, and consequently the number of bubbles in a unit volume $L^3$ fell within the range $N_{\text{tot}}=54-318$. 

As we expect, the average lifetime $\bar{T}$ and the standard deviation $\sigma_{\bar{T}}$ decrease as the transition rate $\beta$ increases. They also decrease as bubble wall speeds $v$ increase.  We also note that the average radius of the bubble at the time of collision $\bar{R}$ is smaller than the radius of bubbles which have nucleated simultaneously and uniformaly at the time of collision $R_{\text{uni}}$. This was also an expected result since bubbles which have nucleated at later times can only nucleate within small pockets of space that still reside in the false vacuum. This shows that the size of the bubbles at the time of collision is misestimated in the simulations where $R_\text{uni}$ was given as the average radius at first collision \cite{Hindmarsh:2015qjv}. 

Interestingly, when we fit the distribution of bubble sizes at the time of collision $\nu_R(\bar{R}_i)$ as a function of the average radius of bubbles at the time of collision $\bar{R}_i=v\bar{T}_i$, we noticed that decays which yielded the same number of bubbles $N_{\text{tot}}$ have produced the same distributions. This indicates that $\nu_R(\bar{R}_i)$ may only depend on the number of bubbles  $N_{\text{tot}}\propto \frac{\beta^3}{v^3}$.

The mathematical framework presented is useful in producing bubble lifetime distributions   $\nu_{T}(\bar{T}_i)$ for a range of parameters. However, as $\beta$ increases, and the number of bubbles $N_{\text{tot}}$ becomes very large, the computations become very time consuming. But we expect that as the number of bubbles $N_{\text{tot}}$ increases, the estimate that the average radius of the bubble at the time of collision $\bar{R}$ is given by the radius of the bubbles which have nucleated uniformaly and simultaneously at the time of their collision $R_{\text{uni}}\propto N^{-1/3}_{\text{tot}}$ becomes more viable.  

We relied on expressions given in Enqvist  \textit{et al}  \cite{Enqvist:1991xw} to describe the times of nucleation $t_n$ and the fuction $h(t)$ which gives the fraction of space that resides the metastable phase as a function of time. By comparing the resultant distributions with distributions resulting from stacking up 1000 simulations of randomly generated periodic Universes with the same number bubbles per unit volume $n_\text{tot}$, we found that the distributions were of similar shapes and their average values were roughly same. This shows that the mathematical framework yields a good estimate of the distribution of bubble lifetimes. 

We hope to find signatures of such cosmological phase transitions by dectecting resultant gravitational waves in the future. The ESA is planning to launch a laser interferometer into space in the late 2030s under the LISA project \cite{Caprini:2015zlo,Caprini:2019egz}. This will enable us to probe frequencies typical of such cosmological transitions which we expect to have occured in the early Universe.
\subsection*{Acknowledgments}
\vspace{-3mm}
I would like to thank Mark Hindmarsh and Apostolos Pilaftsis for insightful comments. The work of Mulham Hijazi is supported by UKSACB.
\                                         

%\newpage

\end{document}